\let\emph\textit 
\documentclass[conference,a4paper]{APSIPA2020}
\usepackage{multirow}
\pdfoutput=1
\usepackage{graphicx}
\usepackage{amsmath}
\usepackage[psamsfonts]{amssymb}
\usepackage{amsxtra}
\usepackage{threeparttable}

\usepackage{cite}
\usepackage{array}
\usepackage{booktabs}
\usepackage{CJKutf8}
\usepackage{placeins}
\usepackage{subfigure}
\usepackage{url}
\graphicspath{{Figures/}}

\begin{document}
	\begin{CJK}{UTF8}{gbsn}
		\title{Online Speaker Adaptation for WaveNet-based Neural Vocoders}

		\author{%
	\authorblockN{Qiuchen Huang, Yang Ai, Zhenhua~Ling
	}
    \authorblockA{
        National Engineering Laboratory for Speech and Language Information Processing,\\
        University of Science and Technology of China, Hefei, China \\
        E-mail: \{qchuang, ay8067\}@mail.ustc.edu.cn, zhling@ustc.edu.cn \\
        \\
        }

       }

\maketitle
\thispagestyle{empty}

\begin{abstract}
In this paper, we propose an online speaker adaptation method for WaveNet-based neural vocoders in order to improve their performance on speaker-independent waveform generation. 
In this method, a speaker encoder is first constructed using a large speaker-verification dataset which can extract a speaker embedding vector from an utterance pronounced by an arbitrary speaker.
At the training stage, a speaker-aware WaveNet vocoder is then built using a multi-speaker dataset which adopts both acoustic feature sequences and speaker embedding vectors as conditions.
At the generation stage, we first feed the acoustic feature sequence from a test speaker into the speaker encoder to obtain the speaker embedding vector of the utterance.  Then, both the speaker embedding vector and acoustic features pass the speaker-aware WaveNet vocoder to reconstruct speech waveforms.
Experimental results demonstrate that our method can achieve a better objective and subjective performance on reconstructing waveforms of unseen speakers than the conventional speaker-independent WaveNet vocoder.

\end{abstract}


\section{Introduction}
In recent years, speech synthesis has become an essential technique for intelligent speech applications, such as audiobook, customer service, speech translation, etc.
At present, speech synthesis also faces more and more challenges, such as high quality, high efficiency, and better generalization ability toward multi-speakers.

Statistical parametric speech synthesis (SPSS) is one of the mainstream speech synthesis approaches, which is achieved by acoustic modeling and vocoder-based waveform generation. It has advantages of smoothness, flexibility and coherence.
Acoustic models predict acoustic features from input linguistic features and can be built based on hidden Markov models (HMM)\cite{tokuda2013speech}, neural networks or other deep learning methods.
Then, vocoders reconstruct speech waveforms from the predicted acoustic features.
Traditional vocoders usually adopt the source-filter signal processing model, i.e., passing a spectrally flat excitation (impulse train or noise) through a linear vocal tract filter, to reconstruct speech waveforms.
Represented by STRAIGHT \cite{kawahara1999restructuring} and WORLD \cite{morise2016world}, these vocoders are convenient and practical but have some deficiencies.
For example, the process of real speech production contains nonlinear effects, which can not be reflected by linear filtering. In addition, there is the loss of spectral details and phase information in these vocoders.

Recently, deep learning models have been widely applied to various signal processing tasks. 
In the filed of speech synthesis, vocoders based on neural networks have also been studied. 
WaveNet \cite{vanwavenet}, a non-linear autoregressive waveform generation model, has been proposed and 
WaveNet-based neural vocoders \cite{tamamori2017speaker} outperformed traditional vocoders on the naturalness of generated speech. 
Some variants, including WaveRNN \cite{kalchbrenner2018efficient}, FloWaveNet \cite{kim2018flowavenet}, ClariNet \cite{ping2018clarinet} and WaveGlow \cite{prenger2019waveglow}, 
have also been proposed to improve the performance and efficiency of WaveNet vocoders. 
However, all of the above neural vocoders rely on speaker-dependent model training.
For some applications such as personalized and expressive speech synthesis, the training data of a target speaker is usually limited.
Besides, separate models need to be stored for different speakers which increases the footprint of speech synthesis systems when adding speakers.

\begin{figure*}
	\centering
	  \includegraphics[width=0.85\linewidth]{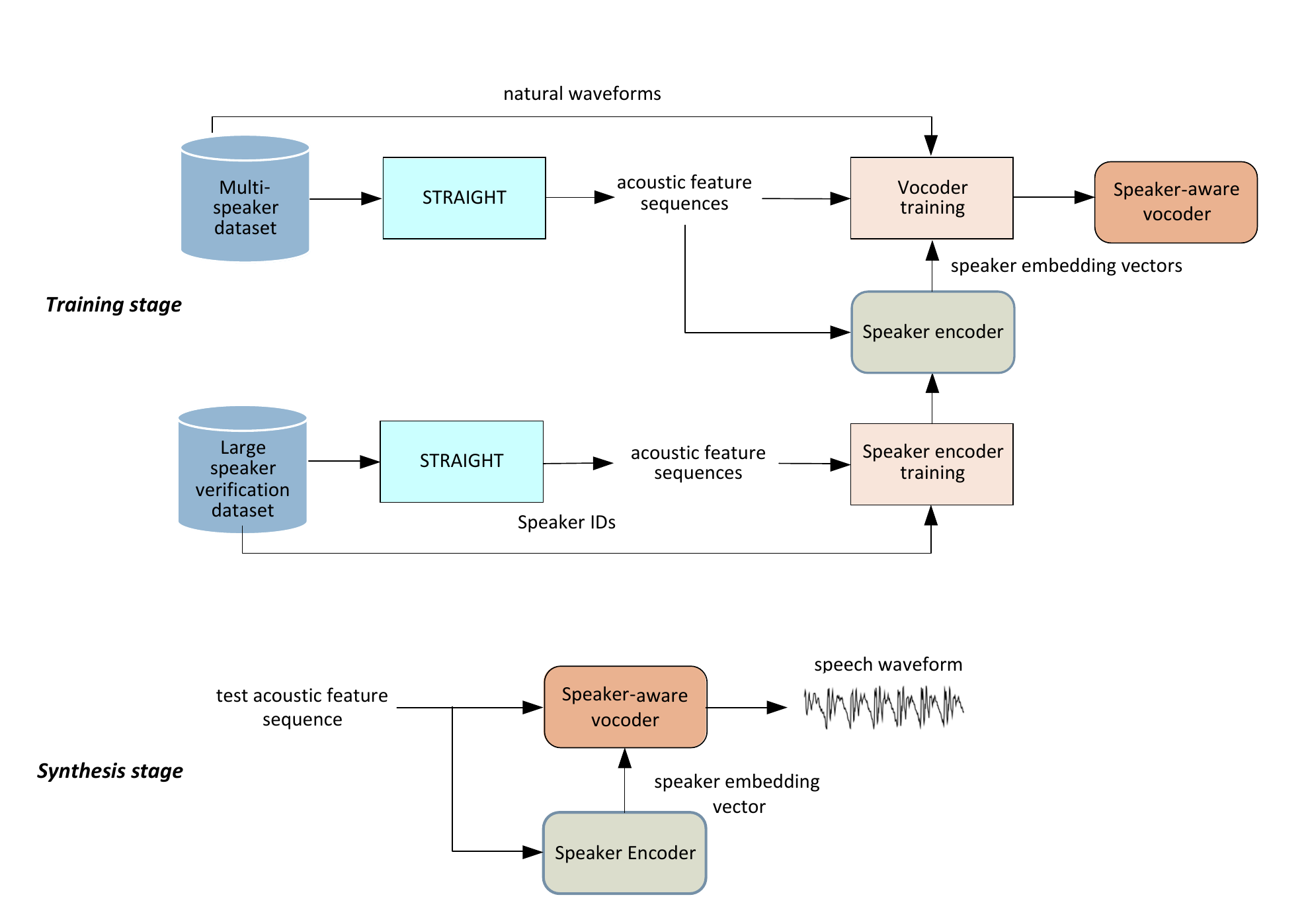}
	\caption{The training and synthesis procedures of our proposed model.}
	\label{fig:structure}
\end{figure*}

The methods of acoustic model adaptation  have been well-studied in traditional SPSS \cite{yamagishi2009analysis}. 
For building neural vocoders, speaker-dependent training methods have also been proposed \cite{liu2018wavenet,wu2019quasi} to avoid the demand for large speaker-dependent training datasets.
Liu et al. \cite{liu2018wavenet} initialized the WaveNet model with a multi-speaker corpus and then fine-tuned it with the small amount of data from the target speaker. 
In this method, speaker embedding vectors were learnt simultaneously with WaveNet parameters at the initialization stage. 
Besides, a quasi-periodic WaveNet vocoder (QPNet) was also proposed\cite{wu2019quasi}, whose dilated convolution structure is adjusted to the  fundamental frequency to enhance pitch controllibiliy for better speaker adaptation. 

These speaker adaptation methods of neural vocoders still require a certain amount of adaptation data and an extra adaption process for each target speaker. 
Thus, they can not achieve fully speaker-independent generation of speech waveforms which is the advantage of traditional source-filter-based vocoders \cite{kawahara1999restructuring,morise2016world}.
Aiming at avoiding this deficiency, the methods of building speaker-independent neural vocoders have been studied using WaveNet \cite{hayashi2017investigation} and WaveRNN \cite{kalchbrenner2018efficient} 
or combining speech production mechanisms \cite{juvela2018speaker}.
Speaker-independent WaveNet vocoder \cite{hayashi2017investigation} used a multi-speaker corpus to train a conditional WaveNet directly.  
However, its performance degraded significantly on unseen speakers comparing with speaker-dependent counterpart \cite{tamamori2017speaker}. 
The reason is that we are unable to cover the voice characteristics of all possible unseen speakers with the training set of vocoders, which makes the built vocoder models prone to overfit to the  speaker characteristics in the training set.

On the other hand, the techniques of representing speaker-specific information using speaker identity embeddings have also been investigated. 
There are mainly three types of speaker identity embeddings: a speaker-code vector (e.g., one-hot vector), an acoustic-driven vector extracted using external models such as i-vector \cite{wu2015study} or d-vector \cite{doddipatla2017speaker,variani2014deep}, and an acoustic-driven vector based on encoders jointly trained with acoustic models \cite{snyder2018x,variani2014deep}.
These speaker identity embeddings have been widely used in speaker identification  and speech recognition tasks \cite{abdel2013fast,heigold2016end,wan2018generalized,zhang2019fully}. 
They have also achieved good performance on building acoustic models for multi-speaker speech synthesis \cite{doddipatla2017speaker,wu2015study,jia2018transfer,gibiansky2017deep,ping2018deep}.

In this paper, 
we integrate d-vectors into neural vocoders and propose an online speaker adaptation method in order to  improve the performance of speaker-independent WaveNet vocoders when dealing with unseen  speakers.
First, a speaker encoder is trained using a large speaker-verificatioin dataset. 
For each utterance in the multi-speaker vocoder training set, a speaker embedding vector, i.e., d-vector, is extracted using the built speaker encoder.
Then, these speaker embedding vectors are utilized as auxiliary features to train a speaker-aware WaveNet vocoder. 
At the generation stage, we send the acoustic feature sequence of a test utterance into the speaker encoder to extract its speaker embedding vector, 
which is combined with acoustic features and passed through the speaker-aware WaveNet vocoder for waveform reconstruction.
Experiment results show that our proposed method can synthesize speech with better objective and subjective quality than the traditional speaker-independent WaveNet vocoder.

Our paper is organized as follows. Section II introduces the details of our proposed method. 
Section III describes the vocoders we built for comparison and the evaluation results. 
Section IV is the conclusion.

\section{Methods}
As shown in Fig. 1, our proposed model is composed of two separately trained neural networks: a recurrent speaker encoder that computes the speaker embedding vector from the input acoustic features of each utterance, and an auto-regressive WaveNet vocoder that utilizes the concatenation of speaker embedding and acoustic features as condition for waveform reconstruction.

\subsection{Speaker Encoder}
The speaker encoder is employed to extract speaker embedding vectors from the acoustic features of target speakers for assisting the vocoder network to generate waveforms. 
The extracted speaker embedding vectors are expected to reflect the speaker characteristics of input acoustic features rather than text contents or background noise.

For building the speaker encoder, we refer to previous study \cite{wan2018generalized} which proposed an efficient and accurate text-independent speaker verification model based on the generalized end-to-end (GE2E) loss.
This model mapped the acoustic features of an utterance to a fixed-dimensional speaker embedding vector, known as d-vector \cite{doddipatla2017speaker,variani2014deep}.
By optimizing the GE2E loss, the d-vectors of training utterances from the same speaker achieved high cosine similarity, while those of utterances from different speakers became far apart in the embedding space.
 
In our implementation, the speaker encoder is a 3-layer LSTM network, including 768 units with projection size 256. 
The embedding vector (d-vector) is defined as the network output at the last frame and its dimension is the same as the projection size of LSTM network. 
At the inference stage of processing each utterance, we apply a sliding window of fixed 160 frames with 50\% overlap. 
The d-vector of each window is fist computed and the final utterance-wise d-vector is generated by averaging the window-wise ones with L2 normalization. 
Different from previous work for speaker verification \cite{wan2018generalized}, 
the acoustic features used to train the speaker encoder here are consistent with the input features of our WaveNet vocoder, including 40-dimensional mel-cepstra, an energy, an F0 and a voiced/unvoiced (V/UV) flag for each frame.
STRAIGHT \cite{kawahara1999restructuring} is used for natural acoustic features extraction. The window size is 400 samples (25ms) and the window shift is 80 samples (5ms).

\subsection{Speaker-Aware WaveNet Vocoder}
WaveNet \cite{vanwavenet} is a deep autoregression-based convolutional neural network which can directly generate high-fidelity audio signal sample-by-sample. 
WaveNet vocoder \cite{tamamori2017speaker} models the joint distribution of waveform samples given auxiliary acoustic features which is factorized as a product of conditional probabilities as
\begin{equation}\label{key}
p(\boldsymbol x\mid \boldsymbol h )=\prod_{t=1}^{T}p(x_{t}\mid x_{1},...,x_{t-1},\boldsymbol h),
\end{equation}
where $x_{t}$ is the waveform value at the $t$-th sample, $T$ is the waveform length, and the condition $\boldsymbol h$ is the sequence of acoustic features.
For modeling the conditional probabilities, WaveNet employs a stack of dilated causal convolution layers. 
The history waveforms and condition features pass through these convolution layers with gated activation functions, and predict the posterior probability of current waveform sample with $\mu$-law quantization \cite{recommendation1988pulse}
using a softmax output layer. 

In our proposed method, a speaker-aware WaveNet vocoder is built, which means that the condition $\boldsymbol h$  contains not only the acoustic features of input utterance but also the speaker embedding vector extracted from these acoustic features by the speaker encoder.
By introducing speaker embedding vectors, we expect to improve the speaker-independency and the speaker-generalization ability of the WaveNet model.
In our implementation, the speaker embedding vector is concatenated with the acoustic features at each frame. Then, they pass through a conditional network consisting of a $ 1\times1  $ convolution layer, a stack of 4 dilated convolution layers and an upsampling layer before acting as the sample-wise local conditions of the WaveNet model. 

As shown in Fig.~1~a), a multi-speaker dataset is adopted to train the speaker-aware WaveNet vocoder.
Acoustic features are first extracted from all training utterances by STRAIGHT \cite{kawahara1999restructuring}. 
Then, a speaker embedding vector is extracted from each utterance using the built speaker encoder. 
It is expected that the speaker embedding vectors can capture speaker-related information in the  training set. 
Finally, the parameters of the speaker-aware WaveNet model are estimated under cross-entropy criterion using waveforms, acoustic features together with speaker embedding vectors of the training set.

At the synthesis stage, given the acoustic features of a test utterance, we first extract its speaker embedding vector as shown in Fig.~1~b).
Then, the extracted speaker embedding vector are concatenated with the acoustic features for waveform construction.
Here, the online adaptation of WaveNet vocoder is achieved because the speaker embedding vector is calculated for each input utterance from an arbitrary speaker and it is not necessary to conduct model adaptation offline using pre-collected data.

\section{Experiments}

\subsection{Datasets}
The VCTK corpus \cite{veaux2016superseded} was adopted to build vocoders in our experiments.
This dataset was downsampled to 16kHz and contained 44-hours utterances recorded from 109 native speakers of Engilish with various accents. 
The corpus was split into three disjoint sets for experiments.
34,977 utterances from 99 speakers and 288 utterances from 10 unseen speakers were chosen to construct the training set and the test set of vocoders. 
The  remaining 3,028 utterances from the 10 unseen speakers were used as  the offline adaptation set. 

For building the speaker encoder, a dataset with more speakers is necessary in order to deal with diversified speakers. 
Thus, the subsets of Librispeech and Voxceleb1 corpora used by previous study on speaker diarizization \cite{zhang2019fully} was adopted here. 
The train-other-500 subset of Librispeech \cite{panayotov2015librispeech} contained 148,688 utterances from 1,166 speakers, and the dev subset of Voxceleb1 \cite{nagrani2017voxceleb} 
contained over 147,935 utterances from 1,211 speakers. 
And we used the same test set as that of vocoder to evaluate the performance of speaker encoder.

\subsection{Model Construction}
To investigate the effectiveness of our proposed method, we built three types of WaveNet-based vocoders for comparison using the same VCTK training set. 
The configurations of them are described as follow.

\subsubsection{Speaker Independent (SI) Vocoder}
This vocoder was built by training a unified WaveNet model without speaker embeddings  \cite{hayashi2017investigation} and acted as the baseline in our experiments.
The WaveNet configurations were the same as the ones of our proposed models. 

\subsubsection{Offline Speaker Adaptation (SA) Vocoders}
For each speaker in the test set, five speaker-dependent WaveNet vocoders were built by fine-tuning the SI vocoder using 20\%, 40\%, 60\%, 80\%, and 100\% adaptation data of this speaker respectively.   

\subsubsection{Online Speaker Adaptation (OSA) Vocoders}
As shown in Table I, two proposed vocoders were built using the speaker encoders estimated with different training sets. 
For the OSA1 vocoder, we used the Librispeech and Voxceleb1 datasets introduced above to train the speaker encoder. 
For the OSA2 vocoder, we further added 99 speakers in the VCTK training set to train the speaker encoder. 
The outputs of both speaker encoders were 256-dimensional speaker embedding vectors. 
And the built WaveNet model had 4 convolutional blocks. 
Each block had 10 dilated casual convolution layers whose filter width was 2 and dilation coefficients were $\{1,2,\dots,2^9\}$.
In the gated activation units, the number of gate channels was 100. 
The number of residual channels and skip channels was 100 and 256 in the residual architectures respectively. 
The waveform samples were quantized by 8-bit $\mu$-law. 
In the condition network, the 299-dimensional condition first passed through a 1$\times$1 convolution layer and a stack of 4 dilated convolution layers. 
The channels of all the convolutional layers were 80. 
The filter size was 3 and the dilation coefficients were $\{1,2,4,8\}$ for the dilated convolution layers. 
Finally the 80-dimensional output was connected to the gated activation units after upsampling and the upsampling was performed by repeating the output within each frame. 
The training target was to minimize the cross-entropy and an \emph{Adam} optimizer \cite{kingma2014adam} was adopted to update the model parameters.
The initial learning rate was 0.0001 and the learning rate halved every 100000 steps. 
The model was totally trained for 400000 steps. 
Models were trained and evaluated on a single Nvidia 1080Ti GPU.

\begin{table}[!t]
	\centering
	\caption{Speaker verification EERs (\%) of two speaker encoders on unseen speakers.}
	\begin{tabular}{ccc}
		\hline
		\textbf{System} & \textbf{Speaker Encoder Training Datasets} & \textbf{EER($\%$)} \\
        \hline
		OSA1 & Librispeech,Voxceleb1 & 2.96 \\
		
		OSA2 & Librispeech,Voxceleb1,VCTK & 1.07 \\
		\hline
	\end{tabular}%
	\label{tab:addlabel}%
\end{table}%

\begin{table*}[htbp]
	\centering
	\caption{Objective evaluation results on test set. Here, SA stands for the vocoder using 100\% adaptation data of each speaker.}
	\begin{tabular}{cccccc}
		\hline
		\textbf{Model} & \textbf{SNR(dB)} & \textbf{RMSE-LAS(dB)} & \textbf{ MCD(dB)} & \textbf{RMSE-F0(cent)} & \textbf{V/UV error rate(\%)} \\
		\hline
		SI    & 3.05 & 8.60 & 2.03 & 69.88 & 7.71 \\
		
		SA\_20\%    & 3.45 & 8.49 & 2.00 & 62.90 & 6.27 \\
		
		SA\_40\%    & 3.56 & 8.51 & 2.00 & 53.97 & 6.04 \\
		
		SA\_60\%    & 3.69 & 8.47& 2.00 & 49.82 & 5.90 \\
		
		SA\_80\%    & 3.76 & 8.48 & 2.00 & 47.27 & 5.96 \\
	
		SA    & \textbf{3.83} & 8.49 & 2.00 & \textbf{45.97} & \textbf{5.86} \\
	
		OSA1  & 3.29 & \textbf{8.43}  & \textbf{1.96} & 55.28 & 6.23 \\
	
		OSA2 & 3.30 & 8.46 & 1.98 & 51.17 & 6.52 \\
		\hline
	\end{tabular}%
	\label{tab:addlabel}%
\end{table*}%

\subsection{Performance of Speaker Encoders}
To measure the performance of built speaker encoders, we computed the equal error rates (EERs) of speaker verification on the VCTK test set and the results are shown in Table I. 
We enrolled the utterances from the 10 speakers in the VCTK test set. 
For both speaker encoders, the enrolled and verification speakers were unseen at the  training stage. 
EER was calculated by pairing each test utterance with each enrolled speaker. 
From Table I, we can see that the EERs of both vocoders were relatively low which indicated that both speaker encoders can extract speaker-related information from acoustic feature sequences effectively.
Besides, the speaker encoder of OSA2 achieved lower EER than that of OSA1 because the VCTK training set was also utilized to train the speaker encoder of OSA2, which may reduce the mismatch between the training and test data of the encoder.

\begin{table}[htbp]
	\centering
	\caption{Average preference scores (\%) on naturalness and similarityness among systems, where N/P stands for “no preference” and p denotes the p-value of a t-test between two vocoders.}
	\begin{tabular}{c|c c c c c c}
		\hline
		& \textbf{SI} & \textbf{SA} & \textbf{OSA1} & \textbf{OSA2} & \textbf{N/P} & \emph{p} \\
		\hline
		\multicolumn{1}{c|}{\multirow{4}[2]{*}{Naturalness}} & 26.00  & -     & -     &\bfseries{40.83}  & 33.17  & $<$0.01 \\
		& 33.64  & -     & 38.64  & -     & 27.72 & 0.13   \\
		& 21.88  & \textbf{54.22}  & -     & -     & 23.90  & $<$0.01 \\
		& -     & \textbf{35.00}  & -     & 28.06  & 36.94  & $<$0.05 \\
		\hline
		\multicolumn{1}{c|}{\multirow{4}[2]{*}{Similarity }} & 23.50  & -     & -     & 26.83  & 49.66  & 0.50  \\
		& 27.57  & -     & 31.82  & -     & 40.61  & 0.16  \\
		& 15.16  &  \textbf{38.28} & -     & -     & 46.56  & $<$0.01 \\
		& -     & \textbf{26.25}  & -     & 20.69  & 53.06  & $<$0.05 \\
		\hline
	\end{tabular}%
	\label{tab:addlabel}%
\end{table}%

\begin{figure*}
	\centering
	\includegraphics[width=1\linewidth]{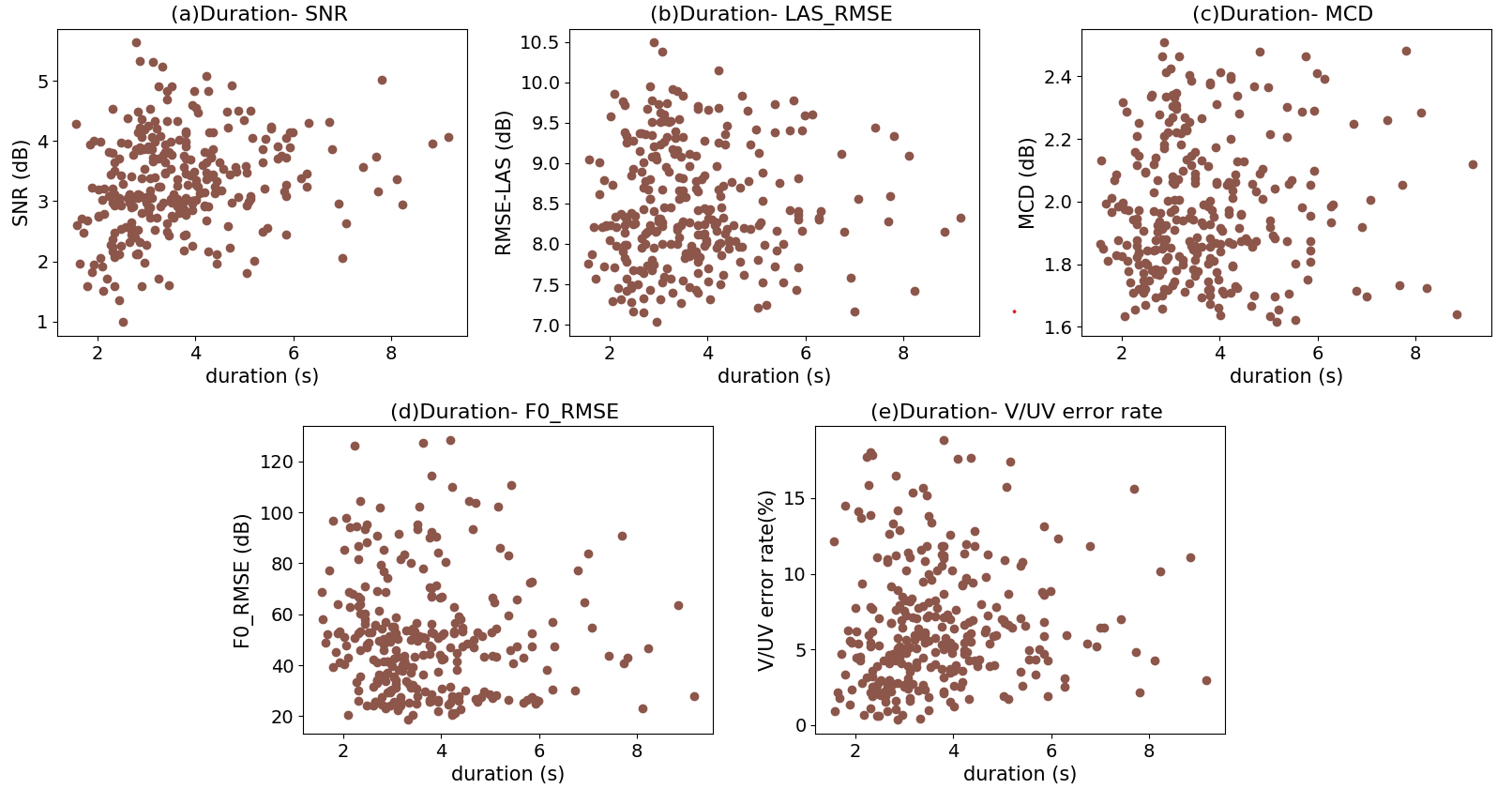}
	\caption{The scatter diagrams between utterance durations and objective evaluation metrics on 288 test utterances.}
	\label{fig:structure}
\end{figure*}

\subsection{Objective Evaluation Results}


Five objective metrics comparing the waveforms generated by vocoders with natural references, including signal-to-noise ratio (SNR), root mean square error of log amplitude spectra (RMSE-LAS),
mel-cepstral distortion (MCD), root mean square error of F0 (RMSE-F0), and voice/unvoiced  error rate (V/UV Error),
were used in our objective evaluation.
The detailed formulae for calculating these metrics can be found in previous studies  \cite{tamamori2017speaker,hayashi2017investigation}.



The test set evaluation results are shown in Table II. 
Regarding with the five SA vocoders, we can see that their performances on SNR, RMSE-F0 and V/UV error rate  were improved when using more adaptation data, while their spectral distortions (i.e., RMSE-LAS and MCD) were almost the same. Comparing the proposed OSA vocoders with SA vocoders, we can see that the spectral distortions of OSA models were comparable with or better than that of SA models, while the SNRs of OSA models were worse than that of SA models, no matter how much adaptation data was used by SA models. The F0-RMSEs of OSA models were close to that of the SA model using 40\% adaptation data, and the V/UV error rate of OSA1 was comparable with that of the SA model using 20\% adaptation data.  Furthermore, although OSA2 achieved lower EER than OSA1 as shown in Table I, its objective performances were not clearly better than OSA1, except on the metric of F0-RMSE.   

\begin{table}[htbp]
	\centering
	\caption{The Pearson's correlation coefficients between utterance durations and evaluation metrics, where $C$ stands for the coefficients and $p$ stands for $p$-value of significance test.}
	    \setlength{\tabcolsep}{4pt}
		\begin{tabular}{cccccc}
		\hline
		& \textbf{SNR} & \textbf{RMSE-LAS} & \textbf{MCD} & \textbf{RMSE-F0} & \textbf{V/UV Error Rate} \\
	    \hline
		C     & 0.23 & 0.01 & 0.06 & -0.08 & 0.09 \\
		p     & $<$0.01 & 0.80 & 0.30 & 0.20 & 0.13 \\
		\hline
	 
	\end{tabular}%
\label{tab:addlabel}%
\end{table}%
\subsection{Subjective Evaluation Results}

For comparing the subjective performance of the speech generated by SI, SA and OSA2 vocoders, four groups of ABX preference tests were conducted on the crowdsourcing platform of Amazon Mechanical Turk with anti-cheating considerations.
In each test, 20 utterances generated by two comparative vocoders were randomly picked out from the test set. Each pair of voice was evaluated in random order. At least 30 English native speakers  were asked to judge which utterance in each pair had better naturalness or  sounded more similar to the natural reference. 
In addition to average preference scores, the p-value of t-test was also calculated to measure the significance of the difference between two comparative vocoders.

The subjective evaluation results are shown in Table III. 
We can see that the OSA2 vocoder achieved better naturalness of reconstructed speech than the SI vocoder significantly (p $<$ 0.01). Meanwhile, we can find that the SA vocoder outperformed both SI and OSA2 vocoders on both naturalness and similarity. The preference score differences between SA and OSA2 were smaller than those between SA and SI. All these results indicate that our proposed method is able to improve the subjective quality of reconstructed waveforms comparing with the traditional speaker-independent WaveNet vocoder. Although OSA1 also achieved higher preference scores than the SI vocoder on both naturalness and similarity, their differences were insignificant (p $>$ 0.05), which implies that the subjective performance of OSA1 was still not as good as OSA2. One possible reason is the mismatch between the training and test data of the speaker encoder used by OSA1, which led to the higher EER and F0-RMSE of OSA1 as shown in Table I and II.

 \subsection{Correlation Analysis between Model Performance and Utterance Duration }
A correlation analysis was conducted to investigate the relationship between the objective performance of the OSA2 vocoder and the duration of test utterances.
288 utterances were randomly selected from the VCTK test set and their durations varied from less than 2 seconds to more than 8 seconds.
These utterances were reconstructed into waveforms using the OSA2 vocoder and the five metrics used in Section III.D were calculated for each utterance.
The scatter diagrams and the Pearson's correlation coefficients between utterance durations and evaluation metrics are shown in Fig. 2 and Table IV respectively.
We can see that though there were no correlations between utterance durations and most other evaluation metrics, the SNR metric has a significant weak correlation relationship to duration. This indicates the performance of our model is influenced somewhat by duration of test utterances and the longer utterance generated is more likely to achieve better quality.

\section{Conclusions}
In this paper, we have proposed an online speaker adaptation method based on a discriminatively-trained speaker encoder and a speaker-aware WaveNet vocoder to improve the performance of traditional speaker-independent neural vocoders. To demonstrate the effectiveness of our proposed model, we also built speaker-independent model (SI) and offline speaker adaptation (SA) vocoders for comparison in our experiments.
Experimental results have demonstrated that our method can achieve lower distortion and better naturalness of reconstructed waveforms than the SI vocoder when dealing with unseen speakers.
Although this paper focuses on WaveNet vocoders, it is also possible to apply our proposed online adaptation method to other neural vocoders, which will be a task of our future work.



\section*{Acknowledgment}
This work was supported by the
National Nature Science Foundation of China (Grant No. 61871358).


\bibliographystyle{IEEEtran}
\bibliography{mybib}

\end{CJK}
\end{document}